\documentclass[11pt]{article}
\usepackage{geometry}
\usepackage{graphicx}
\usepackage{amssymb}
\usepackage{mathtools}
\usepackage{epstopdf}
\usepackage{verbatim}
\usepackage{color}
\usepackage{slashed}
\usepackage{bbold}
\usepackage{fullpage}
\usepackage{authblk}
\usepackage[title]{appendix}
\usepackage[colorlinks, citecolor=red, linkcolor=blue]{hyperref}

\begin{document}

\title{{\Large \textbf{String/M-theories About Our World Are Testable in the
Traditional Physics Way}\thanks{%
This writeup is based on an invited talk at the meeting ``Why Trust a
Theory? Reconsidering Scientific Methodology in Light of Modern Physics",
LMU, Munich, December 2015, and some related talks.}}}
\author{Gordon~L.~Kane \thanks{%
Email:~gkane@umich.edu}}

\affil{\it{Michigan Center for Theoretical Physics (MCTP),}\\ \it{Department
of Physics, University of Michigan}, \\ \it{Ann Arbor, MI 48109 USA}}
\maketitle

\begin{abstract}
Some physicists hope to use string/M-theory to construct a comprehensive
underlying theory of our physical world -- a ``final theory". Can such a
theory be tested?A quantum theory of gravity must be formulated in 10
dimensions, so \textit{obviously }testing it experimentally requires
projecting it onto our 4D world (called ``compactification"). Most string
theorists study theories, including aspects such as AdS/CFT, not phenomena,
and are not much interested in testing theories beyond the Standard Model
about our world. Compactified theories generically have many realistic
features whose necessary presence provides some tests, such as gravity,
Yang-Mills forces like the Standard Model ones, chiral fermions that lead to
parity violation, softly broken supersymmetry, Higgs physics, families,
hierarchical fermion masses and more. All tests of theories in physics have
always depended on assumptions and approximate calculations, and tests of
compactified string/M-theories do too. String phenomenologists have also
formulated some explicit tests for compactified theories. In particular, I
give examples of tests from compactified M-theory (involving Higgs physics,
predictions for superpartners at LHC, electric dipole moments, and more). It
is clear that compactified theories exist that can describe worlds like
ours, and it is clear that even if a multiverse were real it does not
prevent us from finding comprehensive compactified theories like one that
might describe our world. I also discuss what we might mean by a final
theory, what we might want it to explain, and comment briefly on multiverse
issues from the point of view of finding a theory that describes our world.
\end{abstract}

\newpage
\section{Outline}

\begin{itemize}
\item Testing theories in physics -- some generalities.

\item Testing 10-dimensional string/M-theories as underlying theories of 
\textit{our }world obviously \textit{requires }compactification to four
space-time dimensions.

\item Testing of all physics theories requires assumptions and
approximations, hopefully eventually removable ones.

\item Detailed example: existing and coming tests of compactifying M-theory
on manifolds of G$_{2}$ holonomy, \textit{in the fluxless sector}, in order
to describe/explain \textit{our} vacuum.

\item How would we recognize a string/M-theory that explains our world, a
candidate ``final theory"? What should it describe/explain?

\item Comments on multiverse issues from this point of view. Having a large
landscape clearly does not make it difficult to find candidate
string/M-theories to describe our world, contrary to what is often said.

\item Final remarks
\end{itemize}

\section{Long Introduction}

The meeting "Why Trust a Theory" provided an opportunity for me to present
some comments and observations and results that have been accumulating. The
meeting was organized by Richard Dawid, author of \textit{String Theory and
the Scientific Method, }a recent and significant book. Near the end I will
comment on some of Dawid's points.

String/M-theory is a very promising framework for constructing an underlying
theory that incorporates the Standard Models of particle physics and
cosmology, and probably \textit{addresses} all the questions we hope to
understand about the physical universe. Some people hope for such a theory.
A consistent quantum theory of gravity must be formulated in 10 or 11
dimensions. The differences between 10 and 11 D are technical and we can
ignore them here. (Sometimes the theory can be reformulated in more
dimensions but we will ignore that too.) Obviously the theory must be
projected onto four space-time dimensions in order to test it
experimentally. The jargon for such a projection is ``compactification".

Remarkably, many of the compactified string/M-theories generically have the
kinds of properties that characterize the Standard Models of particle
physics and cosmology and their expected extensions! These include gravity,
Yang-Mills gauge theories (such as the color $SU(3)$ and the electroweak $%
SU(2)\times U(1)),$ softly broken supersymmetry, moduli, chiral fermions
(that imply parity violation), families, inflation and more. For some
theorists the presence of such features is sufficient to make them confident
that the string/M-theories will indeed describe our world, and they don't
feel the need to find the particular one that describes our vacuum. For
others, such as myself, the presence of such features stimulates me to
pursue a more complete description and explanation of our vacuum.

The cosmological constant problem(s) (CC) remain major issues, of course. We
will assume that the CC issues are effectively orthogonal to the rest of the
physics in finding a description/explanation of our world. That is, solving
the CC problems will not help us find our underlying theory, and not solving
the CC problems will not make it harder (or impossible) to find our
underlying theory. The evidence we have is consistent with such an
assumption. Ultimately, of course, it will have to be checked.

The value of the CC may be environmental, a point of view attractive to
many. Some physicists have advocated that a number of the constants of the
Standard Model and beyond are environmental. In particular, their values
are then not expected to be calculable by the normal methods of particle
theory such as a compactified string/M-theory. One problem with that
attitude is that apparently not all of the constants are environmental.
Perhaps the strong CP angle, or the proton lifetime, or the top quark mass
are not, and perhaps more. In particular, note that some advocate that the
Higgs boson mass is environmental, but (as described below) we argue that in
the compactified M-theory the HIggs mass (at least the ratio of the Higgs
mass to the Z mass and perhaps the Higgs vev) is calculable. It seems
possible that the strong CP angle and the proton lifetime are also
calculable. If so, it is still necessary to explain why some parameters
seem to have to be in certain somewhat narrow ranges or the world would be
very different. Note that the allowed ranges are quite a bit larger than the
often stated ones - e.g. if the forces are unified one must change their
strengths together, weakening typical arguments (ph/0408169). But the
issue remains that some parameters need to lie in rather narrow ranged, and
need to be understood. In a given vacuum, one can try to calculate the CC. 

Much has been written and said about whether string/M-theories are testable.
Much of what has been said is obviously not serious or interesting. For
example, obviously you do not need to be somewhere to test a theory there.
No knowledgeable person doubts our universe had a big bang, although no one
was there to observe it. There are several very compelling pieces of
evidence or relics. One is the expansion and cooling of the universe, a
second the properties of the cosmic microwave background radiation, and a
third the nucleosynthesis and helium abundance results. Amazingly,
scientists probably have been able to figure out why dinosaurs became
extinct even though almost everyone agrees that no people were alive 65
million years ago to observe the extinction. You don't need to travel at the
speed of light to test that it is a limiting speed.

Is the absence of superpartners at LHC a test of string/M-theory, as some
people have claimed? What if superpartners are found in the 2016 LHC run --
does that confirm string/M-theory? Before a few years ago there were no
reliable calculations of superpartner masses in well-defined theories. An
argument, called \textquotedblleft naturalness", that if supersymmetry
indeed solves the problems it is said to solve then superpartners should not
be much heavier than the Standard Model partners, implies that superpartners
should have been found already at LEP or FNAL, and surely at Run I of LHC.
All predictions up to a few years ago were based on naturalness, rather than
on actual theories, and were wrong. It turns out that in some compactified
string/M-theories one can do fairly good generic calculations of some
superpartner masses, and most of them (but not all) turn out to be quite a
bit heavier than the heaviest Standard Model particles such as the top
quark, or W and Z bosons. Generically scalars (squarks, sleptons, Higgs
sector masses except for the Higgs boson) turn out to be a few tens of TeV
(25-50 TeV), while gauginos (partners of gauge bosons such as gluinos,
photinos, winos, binos) tend to be of order one TeV. I will illustrate the
mechanisms that produce these results later for a compactified M-theory.

Thus actual compactified string/M-theories generically predict that
superpartners \textit{should not} have been found at LHC Run I. Typically
some lighter gauginos do lie in the mass region accessible at Run II. It
seems odd to call the results of good theories ``unnatural", but the
historical (mis-)use of natural has led to that situation.

Interestingly, many string theorists who work on gravity, black holes,
AdS/CFT, amplitudes, and so on do not know the techniques to study
compactified string/M-theories, and their comments may not be useful. Older
theorists may remember. Much of what is written on the subject of testing
string theory does not take into account the need for compactification,
particularly in blogs and some popular books, which is quite misleading.
That's often also the case for what supposed experts say -- in 1999 a well
known string theorist said at a conference \textquotedblleft string
theorists have temporarily given up trying to make contact with the real
world", and clearly that temporary period has not ended. Sadly, string
theory conferences have few talks about compactified string/M-theories, and
the few that might be there are mainly technical ones that do not make
contact with experiments. String theorists seldom read papers about, or have
seminars at their universities about, compactified string/M-theories that
connect to physics beyond the Standard Model.

But I want to argue that string/M-theory's potential to provide a
comprehensive underlying theory of our world is too great to ignore it.
String/M-theory is too important to be left to string theorists.

Before we turn to actual compactified theories, let's look a little more at
the meaning of testing theories. In what sense is \textquotedblleft F=ma"
testable? It's a claim about the relation between forces and particle
properties and behavior. It might have not been correct. It can be tested
for any particular force, but \textit{not }in general. The situation is
similar for the Schr\"{o}dinger equation. For a given Hamiltonian one can
calculate the ground state of a system and energy levels, and make
predictions. Without a particular Hamiltonian, no tests. What do we mean
when we ask to \textquotedblleft test string theory"? The situation for
string/M-theories is actually quite similar to F=ma. One can test
predictions of particular compactified string theories, but the notion of
\textquotedblleft testing string theory" doesn't seem to have any meaning.
If you see something about \textquotedblleft testing string theory", beware.

Quantum theory has some general properties, basically superposition, that
don't depend on the Schr\"{o}dinger equation (or equivalent). Similarly,
quantum field theory has a few general tests that don't depend on choosing a
force, such as that all electrons are identical (because they are all quanta
of the electron field), or the connection between spin and statistics. Might
string/M-theory have some general tests? Possibly calculating black hole
entropy, but I don't want to discuss that here, because it is not a test
connected to data. Otherwise, there don't seem to be any general tests.
Compactified string/M-theories generically give 4D quantum field theories,
and they also imply particular Yang-Mills forces and a set of massless
zero-modes or particles, so it seems unlikely they will have any general
tests.

In all areas of physics normally one specifies the \textquotedblleft theory"
by giving the Lagrangian. It's important to recognize \textit{that physical
systems are described not by the Lagrangian but by the solutions to the
resulting equations}. Similarly, if string/M-theory is the right framework,
our world will be described by a compactified theory, the projection onto 4D
of the 10/11D theory, the metastable (or stable) ground state, called our
\textquotedblleft vacuum".

One also should recognize that studying the resulting predictions is how
physics has always proceeded. All tests of theories have always depended on
assumptions -- from Galileo's using inclined planes to slow falling balls so
they could be timed, or assuming air resistance could be neglected in order
to get a general theory of motion, to assuming what corner of string theory
and compactification manifold should be tried. Someday there may be a way to
derive what is the correct corner of string theory, or the compactification
manifold, but most likely we will first find ones that work to
describe/explain our world, and perhaps later find whether they are
inevitable. Similarly, in a given corner and choice of manifold we may first
write a generic superpotential and K\"{a}hler potential and gauge kinetic
function and use them to calculate testable predictions. Perhaps soon
someone can calculate the K\"{a}hler potential or gauge kinetic function to
a sufficient approximation that predictions are known to be insensitive to
corrections.

It's very important to understand that the tests are tests of the
compactified theory, but they do depend on and are affected by the full
10/11D theory in many ways. \textit{The curled-up dimensions contain
information about our world} -- the forces, the particles and their masses,
the symmetries, the dark matter, the superpartners, electric dipole moments,
and more. There are relations between observables. The way the small
dimensions curl up tells us a great deal about the world.

Just as a Lagrangian has many solutions (e.g. elliptical planetary orbits
for a solar system), so the string/M-theory framework will have several or
many viable compactifications, many solutions. This is usually what is meant
by \textquotedblleft landscape". The important result to emphasize here is
that the presence of large numbers of solutions is not necessarily an
obstacle to finding the solution or set of solutions that describe our
world, our vacuum. (See also the talk of Fernando Quevedo at this meeting.)
That is clear because already people have found, using a few simple guiding
ideas, a number of compactified theories that are like our world. Some
generic features were described above, and the detailed compactified
M-theory described below will illustrate this better. \textit{It is not
premature to look for our vacuum.}

\section{What might we mean by ``final theory"? How would we recognize it?}

In each vacuum perhaps all important observables would be calculable (with
enough grad students and postdocs encouraged and supported to work in this
area). What would we need to understand and calculate to think we had an
underlying theory that was a strong candidate for a \textquotedblleft final
theory"? We don't really want to calculate hundreds of QCD and electroweak
predictions, or beyond the Standard Model ones. We probably don't need an
accurate calculation of the up quark mass, since at its MeV value there may
be large corrections from gravitational and other corrections, but it is
very important to derive $m_{up}<m_{down},$ and that $m_{up}$ is not too
large. It's interesting and fun to make such a list. Here is a good start.

\vspace{0.3cm} $\checkmark $ What is light?

\begin{itemize}
\item What are we made of? Why quarks and leptons?

\item Why are there protons and nuclei and atoms? Why $SU(3)\times
SU(2)\times U(1)?$

\item Are the forces unified in form and strength?

\item Why are quark and charged lepton masses hierarchical? Why is the down
quark heavier than the up quark?

\item Why are neutrino masses small and probably not hierarchical?

\item Is nature supersymmetric near the electroweak scale?

\item How is supersymmetry broken?

\item How is the hierarchy problem solved? Hierarchy stabilized? Size of
hierarchy?

\item How is the $\mu $ hierarchy solved? What is the value of $\mu ?$

\item Why is there a matter asymmetry?

\item What is the dark matter? Ratio of baryons to dark matter?

\item Are protons stable on the scale of the lifetime of the universe?

\item Quantum theory of gravity?

\item What is an electron?
\end{itemize}

$\vartriangleright $ Why families? Why 3? \vspace{0.3cm}\newline
\indent$\vartriangleright $ What is the inflaton? Why is the universe old
and cold and dark? \vspace{0.3cm}\newline
\indent$\Diamond $ Which corner of string/M-theory? Are some equivalent? 
\vspace{0.3cm}\newline
\indent$\Diamond $ Why three large dimensions? \vspace{0.3cm}\newline
\indent$\Diamond $ Why is there a universe? Are there more populated
universes? \vspace{0.3cm}\newline
\indent$\Diamond $ Are the rules of quantum theory inevitable? \vspace{0.3cm}%
\newline
\indent$\Diamond $ Are the underlying laws of nature (forces, particles,
etc) inevitable? \vspace{0.3cm}\newline
\indent$\Diamond $ CC problems? \vspace{0.3cm}\newline

The first question, what is light, is answered -- if there is an
electrically charged particle in a world described by quantum theory, the
phase invariance requires a field that is the electromagnetic field, so it
has a check, $\checkmark$. The next set, with the bullet, are all addressed
in the compactified M-theory; some are answered. They are all addressed
simultaneously. The next two, with the $\vartriangleright$, are probably
also addressed in compactified string/M-theory. The last six are still not
addressed in any systematic way, though there is some work on them.

The list is presented somewhat technically, but the idea is probably clear
to most readers. Other readers might have a somewhat different list. I'd be
glad to have suggestions. The most important point is that compactified
string/M-theories do address most of the questions already, and will do
better as understanding improves.

The compactified M-theory described below assumed that the compactification
was to the gauge-matter content of the $SU(5)$ MSSM, the minimal
supersymmetric Standard Model. Other choices could have been made, such as $%
SO(10),E_{6},E_{8}.$ So far there is no principle to fix that content. There
are probably only a small number of motivated choices.

\section{Three new physics aspects:}

In compactified theories three things emerge that are quite important and
may not be familiar.

\begin{itemize}
\item The \textquotedblleft \textit{gravitino" }is a spin 3/2 superpartner
of the (spin 2) graviton. When supersymmetry is broken, the gravitino gets
mass via spontaneous symmetry breaking. The resulting mass sets the scale
for the superpartner masses and associated phenomena. For the compactified
M-theory on a G$_{2}$ manifold the gravitino mass is of order 50 TeV. That
is the scale of the soft-breaking Lagangian terms, and thus of the scalars
and trilinear couplings, as well as $M_{Hu}$ and $M_{Hd}$. It also can
contribute to the dark matter mass. Two different mechanisms (described
below) lead to suppressions of the gaugino masses from 50 TeV to $\sim$ 1
TeV, and the Higgs boson.

\item The second is \textquotedblleft \textit{moduli}", which have many
physics effects, including leading to a \textquotedblleft non-thermal"
cosmological history. The curled up dimensions of the small space are
described by scalar fields that determine their sizes, shapes, metrics and
orientations. The moduli get vacuum expectation values, like the Higgs field
does. Their vacuum values determine the coupling strengths and masses of the
particles, and they must be \textquotedblleft stabilized" so the laws of
nature will not vary in space and time. The number of moduli is calculable
in string/M-theories (the third Betti number), and is typically of order
tens to even over 200. In compactified M-theory supersymmetry breaking
generates a potential for all moduli, and stabilizes them. The moduli fields
(like all fields) have quanta, unfortunately also called moduli, with
calculable masses fixed by fluctuations around the minimum of the moduli
potential. In general inflation ends when they are not at the minimum, so
they will oscillate, and dominate the energy density of the universe soon
after inflation ends.

The moduli quanta couple to all particles via gravity, so they have a decay
width proportional to the particle mass cubed divided by the Planck mass
squared. Their lifetime is long, but one can show that generically the
lightest eigenvalue of the moduli mass matrix is of order the gravitino
mass, which guarantees they decay before nucleosynthesis and do not disrupt
nucleosynthesis. Their decay introduces lots of entropy, and therefore
washes out all earlier dark matter, matter asymmetry, etc. They then decay
into dark matter and stabilize the matter asymmetry. That the dark matter
and matter asymmetries both arise from the decay of the lightest modulus can
provide an explanation of the ratio of matter to dark matter, though so far
only crude calculations have been done along these lines. When moduli are
ignored the resulting history of the universe after the big bang is a simple
cooling as it expands, dominated by radiation from the end of inflation to
nucleosynthesis, called a thermal history. Compactified string/M-theories
predict instead a \textquotedblleft non-thermal" history, with the universe
matter dominated (the moduli are matter) from the end of inflation to
somewhat before nucleosynthesis. All the above results were derived from the
compactified M-theory before 2012.

Compactified string theories give us quantum field theories in 4D, but they
give us much more. They predict generically a set of forces, the particles
the forces act on, softly broken supersymmetric theories, and the moduli
that dominate cosmological history and whose decay generates the dark matter
and possibly the matter asylmmetry.

\item The third aspect is that because there are often many solutions we
look for \textit{\textquotedblleft generic"} results. We have already used
"generic" several times above. Generic results are probably not a theorem,
or at least not yet proved. They might be avoided in special cases. One has
to work at constructing non-generic examples. Importantly, predictions from
generic analyses are generically \textit{not }subject to qualitative changes
from small input changes. Most importantly, they generically have no
adjustable parameters, and no fine tuning of results. Many predictions of
compactified string/M-theories are generic, and thus powerful tests. When
non-generic K\"{a}hler potentials are used, the tests become model-dependent
and much less powerful.
\end{itemize}

\section{ Compactified M-theory on a G$_{2}$ Manifold (11-7=4)}

Now we turn to looking at one compactified theory. Of course, I use the one
I have worked on. The purpose here is pedagodical, not review or
completeness, so references are not complete, and I apologize to many people
who have done important work similar to what is mentioned. References are
only given so those who want to can begin to trace the work. From 1995 to
2004 there was a set of results that led to establishing the basic framework.

\begin{itemize}
\item In 1995 Witten discovered M-theory.

\item Soon after, Papadopoulos and Townsend (th/9506150) showed explicitly
that compactifying 11D M-theory on a 7D manifold with G$_{2}$ holonomy led
to a 4D quantum field theory with N=1 supersymmetry. Thus the resulting
world is automatically supersymmetric -- that is not an assumption!

\item Acharya (th/9812205) showed that non-Abelian gauge fields were
localized on singular 3-cycles. The 3-cycles can be thought of as ``smaller"
manifolds within the 7D one. Thus the resulting theory automatically has
gauge bosons, photons and Z's and W's, and their gaugino superpartners.

\item Atiyah and Witten (th/0107177) analyzed the dynamics of M-theory on G$%
_{2}$ manifolds with conical singularities and their relations to 4D gauge
theories.

\item Acharya and Witten (th/0109152) showed that chiral fermions were
supported at points with conical singularities. The quarks and leptons of
the Standard Model are chiral fermions, with left-handed and right-handed
ones having different $SU(2)$ and $U(1)$ assignments, and giving the parity
violation of the Standard Model. Thus the compactified M-theory generically
has the quarks and leptons and gauge bosons of the Standard Model, in a
supersymmetric theory.

\item Witten (ph/0201018) showed that the M-theory compactification could be
to an $SU(5)$ MSSM, and solve the doublet-triplet splitting problem. He also
argued that with a generic discrete symmetry the $\mu $ problem would have a
solution, with $\mu =0.$

\item Beasley and Witten (th/0203061) derived the generic K\"{a}hler form.

\item Friedmann and Witten (th/0211269) worked out Newton's constant, the
unification scale, proton decay and other aspects of the compactified
theory. However, in their work supersymmetry was still unbroken and moduli
not stabilized.

\item Lucas and Morris (th/0305078) worked out the generic gauge kinetic
function. With this and the generic K\"ahler form of Beasley and Witten one
had two of the main ingredients needed to calculate predictions.

\item Acharya and Gukov brought together much of this work in a Physics
Reports (th/0409191)
\end{itemize}

To extend previous work, we explicitly made five assumptions:

$\blacktriangleright $ Compactify M-theory on a manifold with G$_{2}$
holonomy, \textsl{in the fluxless sector. }This is well motivated. The
qualitative motivation is that fluxes (the multidimensional analogues of
electromagnetic fields) have dimensions, so are naturally of string scale
size. It is very hard to get TeV physics from such large scales. There are
still few examples of generic TeV mass particles emerging from
compactifications with fluxes. Using the M-theory fluxless sector is robust
(see Acharya referenced above, and recent papers by Halverson and Morrison,
arxiv:1501.05965; 1412.4123), with no leakage issues.

$\blacktriangleright $Compactify to gauge matter group $SU(5)-MSSM.$ We
followed Witten's path here. One could try other groups, e.g. $SU(3)\times
SU(2)\times U(1),SO(10),E6,E8.$ There has been some recent work on the $%
SO(10)$ case, and results do seem to be different (Acharya et al
1502.01727). Someday hopefully there will be a derivation of what manifold
and what gauge-matter group to compactify to, or perhaps a demonstration
that many results are common to all choices that have $SU(3)\times
SU(2)\times U(1)-MSSM.$

$\blacktriangleright $ Use the generic K\"{a}hler potential and generic
gauge kinetic function.

$\blacktriangleright $ Assume the needed singular mathematical manifolds
exist. We have seen that \textit{many} results do not depend on the details
of the manifold (see below for a list). Others do. There has been
considerable mathematical progress recently, such as a Simons Center
semester workshop with a meeting, and proposals for G$_{2}$ focused
activities. There is no known reason to be concerned about whether
appropriate manifolds exist.

$\blacktriangleright $ We assume that cosmological constant issues are not
relevant, in the sense stated earlier, that solving them does not help find
the properties of our vacuum, and not solving them does not prevent finding
our vacuum. Of course, we would like to actually calculate the CC in the
candidate vacuum or understand that it is not calculable.

We started in 2005 to try to construct a full compactification. Since the
LHC was coming, we focused first on moduli stabilization, how supersymmetry
breaking arises, calculating the gravitino mass and the soft-breaking
Lagrangian for the 4D supergravity quantum field theory, which led to Higgs
physics, LHC physics, dark matter, electric dipole moments, etc., leaving
for later quark and lepton masses, inflation, etc. Altogether this work has
led to about 20 papers with about 500 arXiv pages in a decade.

Electric dipole moments are a nice example of how unanticipated results can
emerge - when we examined the phases of the terms in the soft-breaking
Lagrangian all had the same phase at tree level, so it could be rotated away
(as could the phase of $\mu$), so there were no EDMs at the
compactification scale. Then the low scale phase is approximately calculable
from known RGE running, and indeed explains why EDMs are much smaller than
naively expected (0906.2986; 1405.7719), a significant success.

One can write the moduli superpotential (see below). It is a sum of
exponential terms, with exponents having beta functions and gauge kinetic
functions. Because of the axion shift symmetry only non-perturbative terms
are allowed in the superpotential, no constant or polynomial terms. One can
look at early references (Acharya et al, th/0606262, th/0701034,
arxiv:0801.0478, arxiv:0810.3285) to see the resulting terms.

We were able to show that in the M-theory compactification supersymmetry was
spontaneously broken via gaugino and chiral fermion condensation, and
simultaneously the moduli were indeed all stabilized, in a de Sitter vacuum,
unique for a given manifold. We calculated the soft-breaking Lagrangian, and
showed that many solutions had electroweak symmetry breaking via the Higgs
mechanism.

So we have a 4D effective softly-broken supersymmetric quantum field theory.
It's important to emphasize that in the usual \textquotedblleft effective
field theory" the coefficients of all operators are independent, and not
calculable. \textit{Here the coefficients are all related and are all
calculable.} This theory has no adjustable parameters. In practice, some
quantities cannot be calculated very accurately, so they can be allowed to
vary a little when comparing with data.

\section{Some Technical Details}

Here for completeness and for workers in the field we list a few of the most
important formulae, in particular the moduli superpotential, the K\"{a}hler
potential, and the gauge kinetic function. Readers who are not working in
these areas can of course skip the formulae, but might find the words
somewhat interesting.

The moduli superpotential is of the form

\begin{equation}
W=A_{1}e^{ib_{1}f_{1}}+A_{2}e^{ib_{2}f_{2}}.
\end{equation}

The $b$'s are basically beta functions. Precisely, $b_{k}=2\pi /c_{k}$ where
the $c_{k}$ are the dual coxeter numbers of the hidden sector gauge groups. $%
W$ is a sum of such terms. Each term will stabilize all the moduli -- the
gauge kinetic functions are sums of the moduli with integer coefficients
(written below) so expanding the exponentials a potential is generated for
all the moduli. With two (or more) terms one can see in calculations that
the moduli are stabilized in a region where supergravity approximations are
good, while with one term that might not be so. With two we can also get
some semi-analytic results that clarify and help understanding, so we mostly
work with two terms, though some features are checked numerically with more
terms. This is not a \textquotedblleft racetrack" potential; the relative
sign of the terms is fixed by axion stabilization. The generic K\"{a}hler
potential is

\begin{equation}
K=-3\ln (4\pi ^{1/3}V_{7})
\end{equation}

where the 7D volume is

\begin{equation}
V_{7}=\sum_{i=1}^{N}s_{i}^{a_{i}}
\end{equation}

with the condition

\begin{equation}
\sum_{i=1}^{N}a_{i}=7/3.
\end{equation}

The gauge kinetic function is

\begin{equation}
f_{k}=\sum_{i=1}^{N}N_{i}^{k}z_{i}.
\end{equation}

with integer coefficients. The $z_{i}=t_{i}+is_{i}$ are the moduli, with
real parts being the axion fields and imaginary parts the zero modes of the
metric on the 7D manifold; they characterize the size and shape of the
manifold.

Generically two 3D sub-manifolds will not intersect in the 7D space, so no
light fields will be charged under both the Standard Model visible sector
gauge group and any hidden sector gauge group, and therefore supersymmetry
breaking will be gravity mediated in M-theory vacua. This is an example of a
general result for the compactified M-theory, not dependent in any way on
details of the K\"{a}hler potential. It is not automatic in other corners of
string theory, and indeed often does not hold in others. The 11D Planck
scale is

\begin{equation}
M_{11}=\sqrt{\pi }M_{pl}/V_{7}^{1/2}
\end{equation}%
and lies between the unification scale and the 4D Planck scale, which is
related to the absence of fluxes. Acharya and Bobkov have calculated the
cross term in the K\"{a}hler potential between the moduli and the matter
sector; the only results currently sensitive to that are the Higgs mass and
the precise value of the gravitino mass, and it has been included in their
calculation (arxiv:1408.1961).

\section{Main Results, Predictions, and Tests of the Compactified M-theory
So Far, and In Progress}

The results listed here follow from the few discrete assumptions listed
above. Note that the results hold simultaneously. The only dimensionful
parameter is the Planck constant, which is related to Newton's constant G.

\begin{itemize}
\item All moduli are stabilized. Their vacuum expectation values (vevs) are
calculated, and typically $\lesssim \frac{1}{10}M_{Pl}.$ The moduli mass
matrix is calculable and one can find its eigenvalues (th/0701034).

\item The lightest moduli mass matrix eigenvalue has about the same mass as
the gravitino, for general reasons (arxiv:1006.3272).

\item The gravitino mass is calculated approximately to be about 50 TeV,
starting from the Planck scale; see Figure 1 (arxiv:1408.1961).

\item The supersymmetry soft-breaking Lagrangian is calculated at high and
low scales. Scalar masses (squarks, sleptons, $M_{Hu},M_{Hd})$ are heavy,
about equal to the gravitino mass at the compactification scale. RGE running
leads to the third family being significantly lighter than the first two at
the electroweak scale, and $M_{Hu}$ driven to $\sim 1$ TeV there 
(arxiv:0801.0478; 0810.3285).

\item Trilinear masses are calculated to be somewhat heavier than the
gravitino mass.

\item Gaugino masses are always suppressed since the visible sector gauginos
get no contribution from the chiral fermion F-terms. This is completely
general and robust and just follows from the the supergravity calculations.
Since the matter K\"{a}hler potential does not enter, the results are
reliable. The suppression ratio is approximately the ratio of 3-cycle
volumes to the 7D volume (in dimensionless units), so the gluino mass is
about 1.5 TeV ( $\pm 10-15\%).$ The wino mass is about 640 GeV and the bino
is the LSP, with a mass of about 450 GeV.(th/0606262)

\item The hierarchy problem is solved as long as there are about 50 or more
hidden sectors, which is generically true. That is, all solutions have
gravitino masses in the tens of TeV. This is another result that does not
depend on details or the manifold. Technically the number is given by the
3rd Betti number, which Joyce has shown ranges from a few tens to somewhat
over 200 (for non-singular manifolds) (th/0701034).

\item When we set the potential to zero at its minimum by hand (since we do
not solve the CC problem), we find the gravitino mass is in the tens of TeV
region automatically. In other corners of string theory this does not happen.

\item The cosmological history should be non-thermal, with moduli giving a
matter dominated universe from soon after inflation until the lightest
modulus decays somewhat before nucleosynthesis (arxiv:0804.0863).

\item The lightest modulus generates both the matter asymmetry and the dark
matter, and thus also their ratio. Calculations are consistent with the
observed ratio but are very approximate at this stage (arxiv:1108.5178).

\item No approach could be complete without including $\mu $ in the theory.
Witten took the first step to do that (ph/0201018) by exhibiting a generic
matter discrete symmetry that led to $\mu =0.$ But the moduli are
generically charged under that symmetry, so it is broken when they are
stabilized. We have not been able to calculate the resulting value of $\mu $
after moduli stabilization -- it is not clear whether the original discrete
symmetry is broken, or perhaps a new discrete symmetry emerges. We can make
an estimate of $\mu $ because we know that $\mu $ should vanish if either
supersymmetry is unbroken or if the moduli are not stabilized, so $\mu $
will be proportional to a typical moduli vev times the gravitino mass,
implying $\mu \lesssim \frac{1}{10}M_{3/2}$ given the calculated moduli
vevs. Thus $\mu $ will be of order $3$ TeV (arxiv:1102.0556).

\item Axions are stabilized (Acharya et al 1004.5138), giving a solution to
the strong CP problem, and a spectrum of axion masses. The axion decay
constant is allowed to be as high as about $10^{15}$ GeV.

\item We calculated the ratio of the Higgs boson mass to the Z mass, or
equivalently the coefficient $\lambda $ of the $h^{4}$ term in the Higgs
potential (before the LHC data was reported). The calculation is done via
the soft-breaking Lagrangian at the compactification scale, and then the
results are run down to the electroweak scale (so there is no simple formula
for $M_{h}$). One looks for all solutions that have electroweak symmetry
breaking, and calculates the resulting $M_{h}$. Because the scalars are
heavy, the theory is in what is known as the supersymmetry decoupling Higgs
sector, so the Higgs mass is the same for all solutions, $126.4$ GeV.
Because the top mass is not measured precisely, and the RGE equations depend
on the top Yukawa coupling (and $\alpha _{3}$ somewhat less), the running
introduces an error of about $1.2$ GeV purely from Standard Model physics.
In the decoupling sector the Higgs boson properties are close to those of a
Standard Model Higgs, so the branching ratios were predicted to be those of
the Standard Model (to within a few percent, from loops), as is indeed
observed. The Higgs potential is stable, with $\lambda $ never falling below
about $0.1,$ so the vacuum instability is not an interesting question. The
Higgs mechanism occurs because of radiative electroweak symmetry breaking
and is generic.

The size of the Higgs vacuum expectation value is calculable as well, and it
is one of the most fundamental quantities we want to understand. The
mechanism for the Higgs getting a vev is fully understood. It is called
\textquotedblleft radiative electroweak symmetry breaking" (REWSB), because
radiative corrections lead to a Higgs potential with a minimum away from the
origin, giving a non-zero vacuum value. One might think that because the
gravitino mass and the scalars are tens of TeV that the predicted vacuum
value would also be that size, but in fact the corrections lead to a weak
scale value of $M_{Hu}$ of about a TeV near the weak scale, and as just
described $\mu $ is at most a few TeV, so the naive REWSB would give a Higgs
vev of a few TeV, a full order of magnitude smaller than the gravitino mass.
This is an important success of the compactified theory. Further, the EWSB
conditions imply a cancellation occurs. For a certain value of $M_{Hu}$ the
theory would actually give the observed Higgs vev. The value of $M_{Hu}$ is
given by an expression of the form $f_{M_{0}}M_{0}^{2}-f_{A_{0}}A_{0}^{2},$%
where $f_{M_{0}}$ and $f_{A_{0}}$ are fully calculable Standard Model
physics. $M_{0}$ and $A_{0}$ are from the soft-breaking Lagrangian
calculated from the compactification, and are calculated at tree level. $%
M_{0}$ and $A_{0}$ can have loop corrections and K\"{a}hler corrections,
unfortunately, that are large enough so that the degree of cancellation
cannot currently be determined. So the compactified theory might actually
explain numerically as well as conceptually the Higgs vacuum value, but we
don't yet know. This can only work because the compactified theory predicted
large trilinears, about 1.5 times the gravitino mass. That range had not
been previously studied phenomenologically. In any case, the predicted value
of the Higgs vacuum value is within about an order of magnitude of the
observed value, so it is probably qualitatively understood, and not a
mystery.

\item Interestingly, electric dipole moments are calculable. At tree level
the non-perturbative superpotential leads to all soft-breaking terms having
the same phase, which can therefore be rotated away. Similarly, $\mu $ and $%
B $ have the same phase and a Peccei-Quinn rotation removes their phase. So
at the high scale EDM's are approximately zero. When running the down to the
weak scale non-zero EDM's are generated, with the phases entering via the
Yukawa couplings in the trilinears. If those were completely known one could
calculate the low scale EDM's precisely. Because there is still some
model-dependence in the high scale Yukawas, we can only calculate upper
limits on the EDM predictions. These are about a factor $20$ below the
current limits. Thus the compactified M-theory explains the surprisingly
small EDM's, and provides a target for future experiments (arxiv:0985.2986;
1405.7719).

\item There are no flavor or weak CP problems.

\item Combining the electroweak symmetry breaking results and $\mu $, $\tan
\beta \approx 5.$

\item LHC can observe gluinos, the lighter chargino, and the LSP. To see
higgsinos and the scalars (via associated production with gluinos or winos)
one needs a pp collider with energy near 100 TeV. (arxiv:1408.1961)

\item Many results, such as the gravitino mass of 10's of TeV; the heavy
high scale scalars; the suppressed gaugino masses (gluinos, LSP); gravity
mediation; small EDMs; and more are generic and do not depend on details of
the manifolds.

\item Our main current work is on dark matter. As we've discussed above the
physics of the hidden sectors plays several roles. We live on one that is
our visible sector. Others with large gauge groups have couplings that run
fast, and lead to gaugino condensation and associated supersymmetry breaking
at about $10^{14}$ GeV (the scale at which F-terms become non-zero). Others
have small gauge groups so they run slowly, and condense at MeV or GeV or
TeV scales, perhaps giving stable particles at those scales. We can
calculate the relic density of those light particles, which must be done in
a non-thermal universe in compactified M-theory (and probably in
string-theory worlds in general). (arxiv:1502.05406) We are doing this
systematically, expecting to find some dark matter candidates. The bino LSP
will decay into these lighter stable \textquotedblleft wimps". We are
looking at how generic kinetic mixing is in M-theory. While some people have
looked at what they call \textquotedblleft hidden sectors", mostly what they
look at are not actual hidden sectors of a compactified string/M-theory
(they should probably be called hidden valleys to distinguish), and almost
all of them have calculated the relic densities in a thermal history which
is unlikely to be relevant for string/M-theory hidden sectors.
\end{itemize}

\begin{figure}[tbp]
\centering
\makebox[0pt]{\includegraphics[scale=0.7]{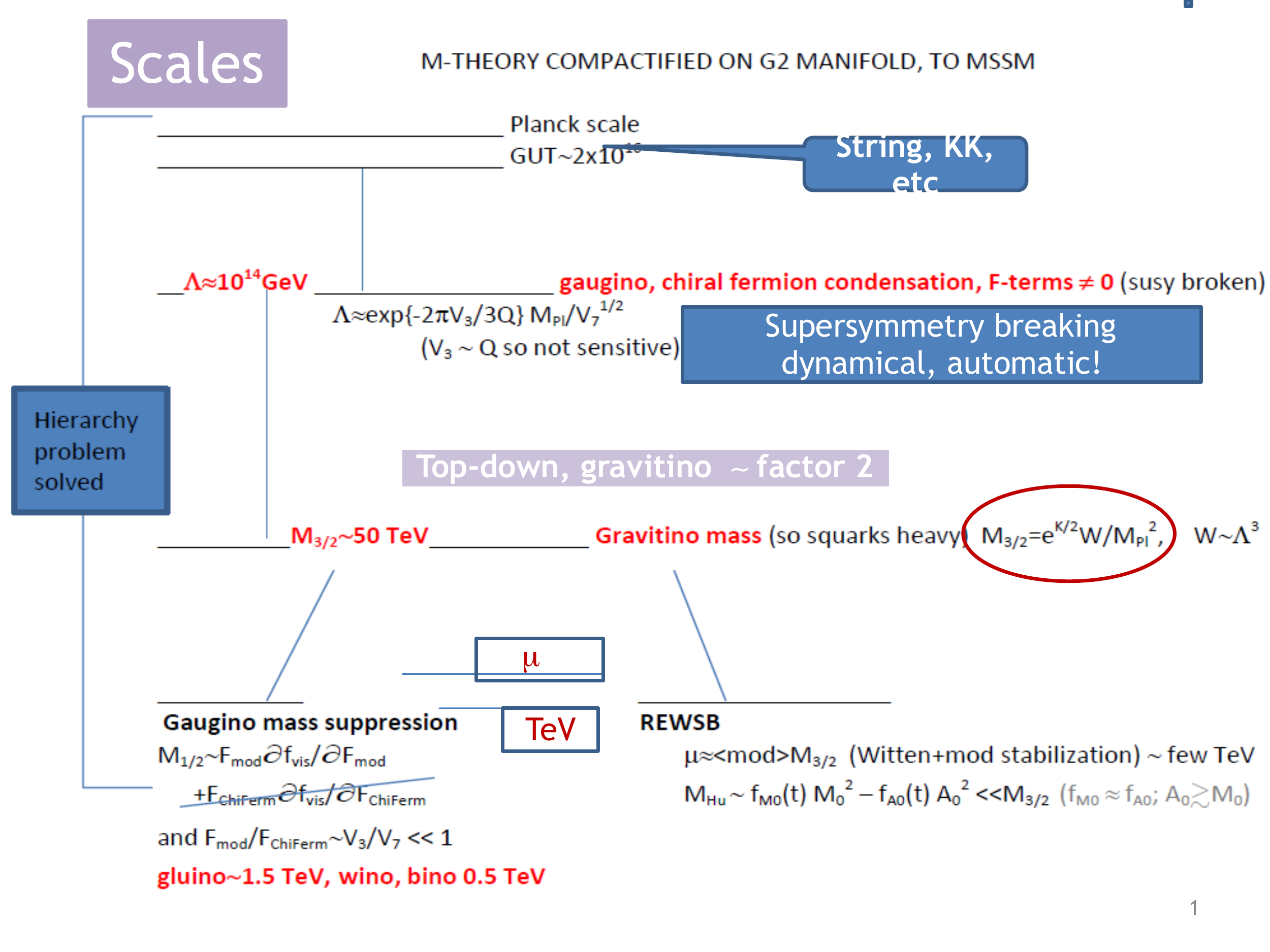}}
\caption{Figure showing the various scales in M-theory compactified on a G$%
_2 $ manifold, with the MSSM as the low energy effective theory.}
\label{Scales.FIG}
\end{figure}

It's interesting to put together the various results on scales to see how
the physics emerges and is connected from the Planck scale to the gaugino
masses and the Higgs mass. This is shown in Figure \ref{Scales.FIG}.

It's worth emphasizing that the gluino (and wino, bino) mass prediction is
not one \textquotedblleft just above current limits", or a tuned
calculation. It is actually very generic and robust, and simple to
understand. F-terms are generated by gaugino (and chiral fermion)
condensation, at about a scale $\Lambda \approx 10^{14}$ GeV, a generic
result of the theory, from the running of the largest hidden sector gauge
groups. The superpotential is then the ratio of this scale to the
compactification scale cubed, and the gravitino mass is the superpotential
times $e^{K/2};$ the latter factor is basically the inverse of the 3-cycle
volume, as shown on the figure. The gaugino suppression is very general,
because of the absence of the chiral fermion contribution to the gaugino
mass since it is a derivative of the visible sector gauge kinetic function
which has no dependence on the chiral fermion F-term (th/0606262). All this
is illustrated in the figure, on the top half and the left side. The Higgs
mass suppression is illustrated on the right side.

The gluino production cross section at LHC is 10-15 fb. Note that this is
significantly smaller (because of the suppressed heavy quark contribution)
than cross sections often quoted in the literature. Because the third family
is lighter due to the RGE running, somewhat over half of the gluino decays
are to third family final states, and the rest to first + second families.
Those signatures make detection more difficult than most LHC studies report
for a given gluino mass, and require larger luminosities than naively
expected according to our background estimates, probably over 40 fb$^{-1}$.
The largest background is top pair production.

There is still of course a lot to do to complete the M-theory
compactification, both the physics and the mathematics. \ It will be very
interesting to try compactifications to other gauge-matter groups in
M-theory, and to pursue other corners of string theory compactifications.

\section{A Comment on Landscape and Multiverse Issues}

There is clearly a landscape of string/M-theory solutions. The question is
whether the resulting universes are viable ones, or too short-lived to have
galaxy formation (with the resulting solar systems, etc. Some recent studies
(Dine, Paban; Mersini, Perry; Greene et al; Shiu et al and others) suggest
most of the landscape does not give viable universes. Even if there is a
large landscape, compactification studies have demonstrated that it is not
hard to find vacua that are good candidates for describing our world and
calculating its properties (except maybe for the CC) -- Quevedo and
collaborators; Nilles and collaborators; Acharya, Kane and collaborators;
Vafa, Heckman. Sometimes it is argued that in a landscape it will be hard to
find our vacuum. It's now clear that finding such candidates is not an
obstacle to finding a final theory for our world.

\section{Final Remarks}

Here is a list of remarks. Some address the topics of the meeting,
\textquotedblleft Why Trust a Theory?", some summarize points made about
compactified string/M-theories, and some focus on the M-theory G$_{2}$
example. I want to emphasize again that this is not a review, and it has
several pedagogical aspects. In particular, there are basically no
references, but some are given (only to arXiv postings) in order to help
people look further into topics they might want to pursue. There is some
overlap among remarks since the point is communication rather than
conciseness.

\begin{itemize}
\item String/M-theory is too important to be left to the string theorists.

\item If you want an underlying theory that is a quantum theory and includes
gravity and the other forces, and the quarks and leptons, a 10/11
dimensional theory with curled up small dimensions is probably the simplest
framework that could incorporate and explain all that you want to understand.

\item The compactified M-theory on a manifold of G$_{2}$ holonomy is a
promising candidate to describe our vacuum -- this at least demonstrates
that it is not premature to look for such theories, even in the presence of
a landscape of solutions.

\item Moduli are generically present in string theories and are inevitable
in M-theory. They imply a non-thermal cosmological history and may explain
the ratio of matter to dark matter.

\item The compactified M-theory anticipated the mass and decay branching
ratios of the Higgs boson. It is the lightest eigenvalue of the Higgs mass
matrix of a two-doublet decoupling supersymmetric Higgs sector that
satisfies the electroweak symmetry breaking conditions. The Higgs potential
does not vanish at any scale, and the universe is metastable. The vacuum
value of the Higgs field is not a mystery.

\item The compactified M-theory predicts that superpartners are too heavy to
have been seen in LHC Run I, but gluinos and winos and the LSP bino can be
seen in Run II at 13 TeV with sufficient luminosity. Backgrounds for
gluinos, mainly from top quark production, imply over 40 fb$^{-1}$are needed
to see the signal.

\item The discovery of the Higgs boson is evidence for supersymmetry. In a
supersymmetric full theory one computes the supersymmetry soft breaking
Lagrangian at the compactification scale. It contains a potential for
scalars. The RGE running down to the TeV scale implies that potential has a
minimum away from the origin. One can calculate accurately the ratio of
Higgs mass to Z mass, and gets 126.4 GeV for the Higgs mass that way. It
does not depend on parameters. The calculation was done before the data, but
the calculation is determined so the answer would be the same whenever it is
done.

\item Compactified string/M-theory imply naturalness predictions are wrong
so superpartners should not have been seen in Run I. Some superpartners can
be seen in LHC Run II with sufficient luminosity (gluinos, winos, binos).
Full testing of the superpartner spectrum will require colliders with total
energy in the 100 TeV region, and sufficient luminosity. The squarks can
only be directly seen at such a facility, in associated production.

\item The LSP will decay into lighter stable particles from hidden sectors.
Such candidates are generic and probably inevitable. We are doing systematic
studies of such dark matter in non-thermal cosmological histories.

\item The statement that estimates of the cosmological constant are off by 10%
$^{120}$ is a red herring, or worse. First, in supersymmetric theories the
value of the potential at its minimum is generically $M_{3/2}^{2}M_{pl}^{2}%
\sim 10^{48}$ GeV$^{4}.$ But this is the potential, and people think in
terms of mass, so take the 4th root, giving $10^{12}GeV$ as the meaningful
value. The Higgs potential and the QCD phase transition give similar values.
Still not so good.

But the QCD strong CP problem requires setting a number naively of order
unity to a value $10^{10}$ times smaller, almost the same! Why is there so
much anguish over one and none over the other? Partly it's because there are
possible solutions to the strong CP problem, such as axions, and good models
that solve it (like compactified string theories), while we don't understand
the CC problem. But partly it is just hype. It is reasonable to think that
the CC problem(s) are orthogonal to the rest of physics, in the sense that
solving the CC problem will probably not help us solve the rest of the
issues about understanding our vacuum, and not solving the CC problem
probably will not prevent us from making progress with understanding our
vacuum.

\item Are string/M-theories falsifiable? Yes, in the same sense as
traditional theories -- one makes predictions from the compactified theories
and tests the predictions. Such predictions are tests of the full underlying
theory plus the gauge-matter compactification group. You do not have to be
there to test whether there was a big bang, and you do not have to be at
Planck scale energies to test string/M-theories formulated there. There are
always relics and implications.

\item Are 10D theories science? Yes, the curled up dimensions after
compactification contain information about the forces and particles -- such
information is not lost, but characterizes the predictions. There are a
number of predictions that test the theories, but not directly the 10D
theories. What could it mean to test a 10D theory? Lots of people talk about
that, but ask them what they mean.

\item What does "empirically testable" mean for string/M-theories?

-- If it means project the the theory onto a 4D world, i.e. compactify, and
find generic predictions, then it is well-defined, has been going on, and is
the way traditional physics has worked. Compactified string/M-theories are
not post-empirical science.

-- If it means including non-generic predictions with additional assumptions
about (say) the K\"{a}hler potential, then it tests the cleverness of the
people doing it as well as testing the theory, and is much less powerful.

-- If it is done for non-compactified theories it does predict a landscape
of solutions, but it cannot address the issue of how many of those solutions
lead to viable worlds that live long enough to contain planets. Even if the
non-viable ones are not universes with people, perhaps they could still have
an effect on the wave function of the universe as part of a superposition.

\item Are string/M-theories forever beyond possibility of testing? For
example, because they are formulated at energies too high to ever reach? No!
Just as no one was at the Big Bang but every knowledgeable scientist knows
it is well tested, by the universe's expansion, by nucleosynthesis and the
helium abundance, and by the embers of the Cosmic Microwave Background, and
additional technical results. Are small extra dimensions a problem? No! The
curled up dimensions contain lots of information that implies testable
predictions for the compactified theories, particularly determining the
forces and particles.

\item Do string/M-theories exist? Can anyone define what they are? This red
herring should not be taken seriously - recall that some of the main
successes of quantum theory in the mid 1920's were achieved before rigorous
definitions of quantum theory were given.

\item Sometimes people say wrongly that the Standard Model offers no path
forward. In fact it tells us to focus on theories beyond the Standard Model
that contain Yang-Mills gauge forces, quarks and leptons with hierarchal
masses, one and only one fermion with a large Yukawa coupling of order gauge
couplings, gravity, etc.

\item Is string/M-theory "the only game in town", Dawid's NAA? It's
important to only consider games that are comprehensive and include all the
issues -- not only gravity, but all the Standard Model forces, existence of
quarks and leptons, dark matter, a cosmic matter asymmetry, electroweak
symmetry breaking, a stable hierarchy of the right size between atoms and
the Planck scale, etc. Compactified string/M-theories are known to address
all these issues, and indeed nothing else is known to do that, not even
close.

\item Dawid's unexpected explanatory power argument is very strong. One of
the best examples is supersymmetry, which was introduced for theoretical
reasons, but turned out to provide possible explanations for the hierarchy
problem, electroweak symmetry breaking, dark matter, matter asymmetry and
more. The unexpected explanatory power of string theory is impressive too --
look at the issues addressed by the compactified M-theory above.

\item One can think of Dawid's book as describing how physicists use theory
assessment at a given stage in order to decide what to work on. It does that
very well. Or, one could think of it as describing how non-physicists might
evaluate the work of physicists (from the general public to philosophers).
Then one could imagine controversy. It holds up well, with much of the
confusion arising from books and articles and blogs that don't understand
the string/M-theories, and that don't understand that only compactified
theories can make contact with our world.
\end{itemize}

\bigskip

\section*{Acknowledgements}

I'm grateful to my collaborators over the past decade on compactified
M-theory, particularly Bobby Acharya and Piyush Kumar as well as Konstantin
Bobkov, Sebastian Ellis, Eric Kuflik, Ran Lu, Jing Shao, Scott Watson, and
Bob Zheng; to David Gross, Brent Nelson, Malcolm Perry, Joe Polchinski and
others for discussions; and to Richard Dawid, Slava Mukhanov, and Gia Dvali
for their hospitality at the meeting.

\bigskip

\bigskip

\bigskip

\bigskip

\bigskip

\bigskip

\bigskip

\bigskip

\bigskip

\bigskip

\bigskip

\bigskip

\bigskip

\bigskip

\bigskip

\bigskip

\end{document}